\documentclass[prb,amsmath,amssymb,superscriptaddress,reprint,twocolumn]{revtex4}
\usepackage[utf8]{inputenc}

\usepackage{hyperref}
\usepackage{dcolumn}
\usepackage{bm}
\usepackage{hyperref}
\usepackage[dvipsnames]{xcolor}
\usepackage{graphicx}
\usepackage[capitalise]{cleveref}
\usepackage{textcomp}
\usepackage{pdfcomment}
\usepackage{bold-extra}
\usepackage{cancel}
\usepackage{tabularx} 
\usepackage{mathrsfs}
\usepackage{makecell}

\usepackage{outlines}
\usepackage{enumitem}
\setenumerate[1]{label=\Roman*.}
\setenumerate[2]{label=\Alph*.}
\setenumerate[3]{label=\roman*.}
\setenumerate[4]{label=\alph*.}

\DeclareUnicodeCharacter{2022}{\,}
\hypersetup{pdftoolbar=true,        
 pdfmenubar=true,        
 pdffitwindow=false,     
 pdfstartview={fitH},    
 pdftitle={},    
 pdfauthor={},     
 pdfsubject={Rev},   
 pdfcreator={},   
 pdfproducer={LaTeX}, 
 pdfkeywords={Version 1.0},
 colorlinks=true,       
 linkcolor=blue,          
 citecolor=blue,        
 filecolor=blue,      
 urlcolor=blue}

\makeatletter
\renewcommand{\@biblabel}[1]{#1. }
\renewcommand{\@dotsep}{500}
\renewcommand{\@pnumwidth}{0em}
\renewcommand{\l@figure}[2]{
\@dottedtocline{1}{1.5em}{2em}{Figure #1}{}\vspace{15pt}}

\usepackage{textcomp}
\usepackage{tikz}

\usepackage[normalem]{ulem}

\begin{document}

\title{Laser injection locking and nanophotonic spectral translation of electro-optic frequency combs}

\author{Roy Zektzer}
 \affiliation{Microsystems and Nanotechnology Division, Physical Measurement Laboratory, National Institute of Standards and Technology, Gaithersburg, Maryland 20899, USA}
\affiliation{Joint Quantum Institute, NIST/University of Maryland, College Park, Maryland 20742, USA}
\author{Ashish Chanana}
\affiliation{Microsystems and Nanotechnology Division, Physical Measurement Laboratory, National Institute of Standards and Technology, Gaithersburg, Maryland 20899, USA}
\author{Xiyuan Lu}
 \affiliation{Microsystems and Nanotechnology Division, Physical Measurement Laboratory, National Institute of Standards and Technology, Gaithersburg, Maryland 20899, USA}
\affiliation{Joint Quantum Institute, NIST/University of Maryland, College Park, Maryland 20742, USA}
\author{David A. Long}\email{david.long@nist.gov}
 \affiliation{Microsystems and Nanotechnology Division, Physical Measurement Laboratory, National Institute of Standards and Technology, Gaithersburg, Maryland 20899, USA}
\author{Kartik Srinivasan}\email{kartik.srinivasan@nist.gov}
 \affiliation{Microsystems and Nanotechnology Division, Physical Measurement Laboratory, National Institute of Standards and Technology, Gaithersburg, Maryland 20899, USA}
\affiliation{Joint Quantum Institute, NIST/University of Maryland, College Park, Maryland 20742, USA}

\date{\today}

\begin{abstract}
High-resolution electro-optic frequency combs (EO combs) consisting of thousands to millions of comb teeth across a bandwidth between 1~GHz to 500~GHz are powerful tools for atomic, molecular, and cavity-based spectroscopy, including in the context of deployable quantum sensors. However, achieving sufficiently high signal-to-noise ratio (SNR) EO combs for use across the broad range of wavelengths required in the aforementioned applications is hindered by the corresponding unavailability of relevant components such as narrow-linewidth lasers, electro-optic phase modulators with adequate optical power handling, and low-noise optical amplifiers. Here, we address the latter two points by showing that optical injection locking of commercial Fabry-Perot (FP) laser diodes can help enable high SNR EO combs. We injection lock the FP laser diode to more than 10$^6$ comb teeth at injected comb powers as low as 1~nW and produce a high SNR replica of the EO comb. In comparison to a commercial semiconductor optical amplifier, injection locking achieves $\approx$100$\times$ greater SNR for the same input power (when $<$1~$\mu$W) and equal SNR for $>35\times$ lower input power. Such low-power injection locking is of particular relevance in conjunction with nanophotonic spectral translation, which extends the range of wavelengths available for EO combs. We show that the usable wavelength range of an EO comb  produced by photo-induced second harmonic generation of an EO comb in a silicon nitride resonator is significantly increased when combined with optical injection locking. Our results demonstrate that optical injection locking provides a versatile and high-performance approach to addressing many different scenarios in which EO comb SNR would be otherwise limited.

\end{abstract}

\maketitle
\section{Introduction}
\label{sec:Intro}
\begin{figure*}[t!]
\centering\includegraphics[width=0.7\linewidth]{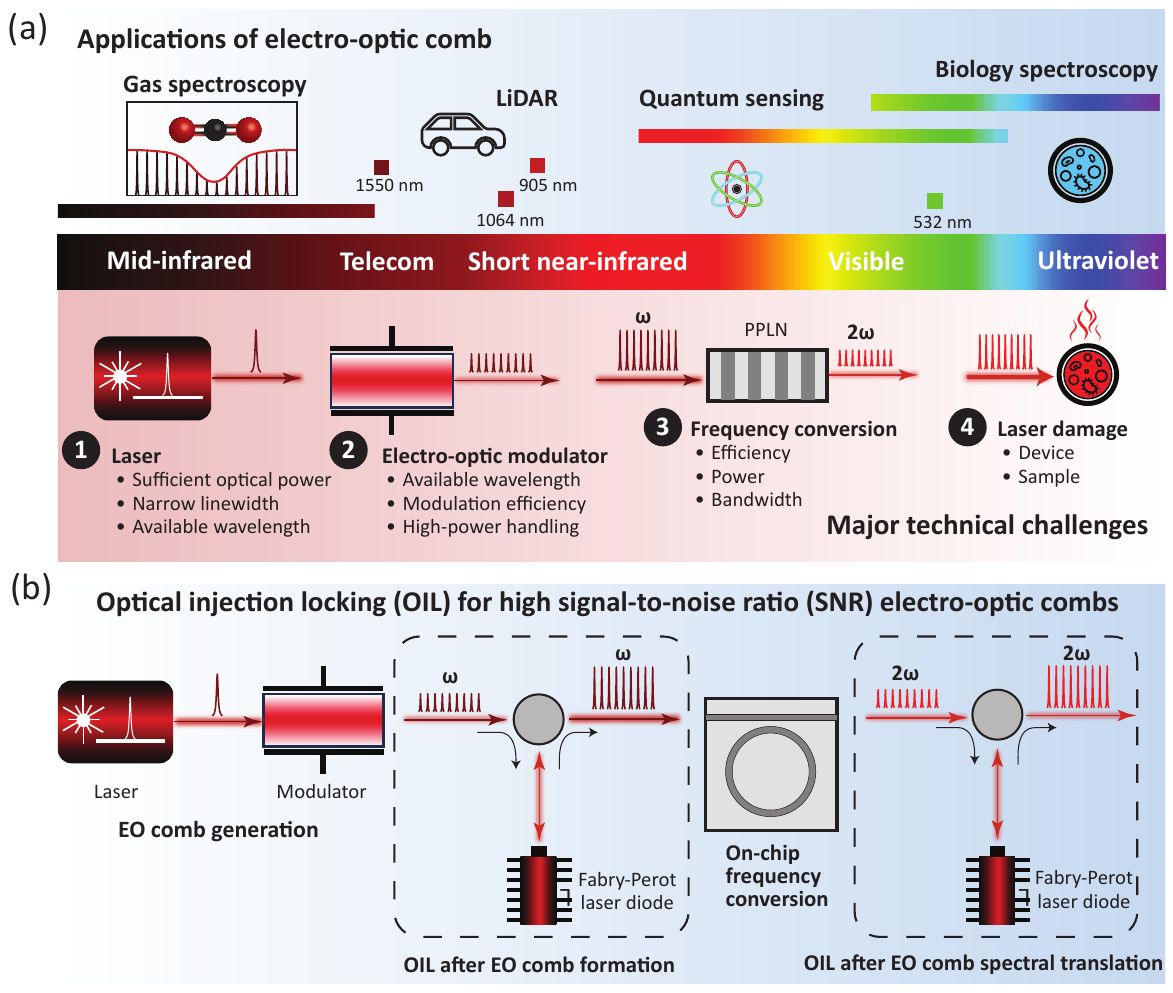}
\caption{ \textbf{Applications and challenges to high signal-to-noise ratio electro-optic (EO) combs, and our proposed solution using optical injection locking.} \textbf{(a)} Current applications of EO combs extend across a wide range of wavelengths from mid-infrared to ultraviolet. These applications include mid-infrared gas spectroscopy~\cite{ns_res_Long2023}, laser detection and ranging (LIDAR)~\cite{he_massively_2023} (automobiles: 1550/905~nm, atmosphere: 1064~nm, and underwater: 532~nm), quantum sensing based on cold atoms and atomic vapors~\cite{rydberg_combPrajapati2024,Dixon2023}, and spectroscopy of biological systems. There are several major technical challenges to high signal-to-noise ratio operation, from the generation of the comb using a narrow-linewidth laser and electro-optic modulator, to frequency conversion of the comb from the laser wavelengths to a target operating wavelength (for example, using periodically poled lithium niobate (PPLN)), as well as laser damage of both devices (e.g., modulator photorefractive damage) and samples. \textbf{(b)} We propose to address these challenges by using optical injection locking (OIL) of Fabry-Perot laser diode(s) to low-power EO combs (left dashed box). This results in a high signal-to-noise ratio EO comb while reducing the optical power at the comb generation step, which limits the requirements on the seed laser and modulator. Moreover, when used after nonlinear spectral translation (right dashed box), it can effectively enhance its conversion efficiency.}
\label{Fig1}
\end{figure*}

\noindent  Optical frequency combs have had a tremendous impact by offering a coherent link between optical and microwave frequencies \cite{Comb_revDiddams2020,Comb_revFortier2019} and enabling applications such as optical atomic clocks~\cite{Atomic_clock_Diddams2001}, laser ranging~\cite{Lidar_Lukashchuk2022}, and spectroscopy~\cite{Sensing_dual_Giorgetta2024}. In recent years, significant efforts have been made to transition combs from the optical table into real-world applications, and even onto chip-scale platforms~\cite{Integrated_combs_Chang2022}. Efforts in this direction include the development of frequency combs based on chip-integrated Kerr nonlinear optics~\cite{microCombrev_Kippenberg2011} and electro-optics~\cite{Int_Eo_Hu2022,Int_EO_Han2024,Int_Eo_Zhang2019,Int_Eo_Zhang2023}. For sensing and spectroscopy applications (Fig.~\ref{Fig1}(a)), electro-optic (EO) combs produced by modulation of continuous-wave lasers~\cite{EO_comb_rev_Parriaux2020,EoCombs_rev_Zhuang2023} are especially interesting as they allow for frequency agile, digitally controlled repetition rates and spans. This enables spectral and temporal resolution that can be tailored to meet the requirements of different materials and applications. For example, EO combs have been used in molecular spectroscopy from the ultraviolet to the mid-infrared spectral regions~\cite{EO_comb_rev_Parriaux2020,EoCombs_rev_Zhuang2023}, allowing for time resolution as fast as 20 ns~\cite{ns_res_Long2023}. The mutual coherence of the comb teeth further enables measurement of ultra-narrow (e.g., $<1$~kHz) spectroscopic features such as those encountered in atomic physics~\cite{long_electro-optic_2019}, allowing application to quantum sensors, as shown in recent demonstrations with Rydberg electrometry~\cite{rydberg_combPrajapati2024, Dixon2023, rydberg_combArtusio2024}.

Considering the variety of applications they impact, there is a strong need for EO combs across a wide range of wavelengths (e.g., ultraviolet to near-infrared) with sufficient power for high signal-to-noise ratio (SNR) measurements. As described below, this presents a significant challenge, particularly at wavelengths outside of the telecommunications (telecom) bands, which contain the most compelling spectroscopic targets but where EO technology is far less mature. Here, we demonstrate that optical injection locking (OIL)~\cite{Liu2020} of a Fabry-P\'erot (FP) laser diode by an EO comb is a versatile tool for addressing SNR challenges in EO comb spectroscopy systems. In contrast to previous approaches where FP laser diode injection locking was seeded by individual comb teeth~\cite{fukushima_optoelectronic_2003,zhang_high-coherence_2024}, here we injection lock a single FP laser diode with the entirety of the EO comb that is comprised of up to two million comb teeth across a span of 2 GHz. High signal-to-noise ratio output combs are created with total injected comb power as low as 1 nW, corresponding to 0.4 pW per comb tooth on average.  In comparison to using a semiconductor optical amplifier (SOA) to boost the EO comb power, we achieve $>100\times$ higher SNR for the same input power of $<$ 1~$\mu$W. Finally, to showcase how our approach naturally dovetails with techniques to extend wavelength access, we combine EO comb spectral translation and OIL. We use doubly-resonant, photo-induced second harmonic generation (SHG) and sum-frequency generation (SFG) in a silicon nitride (Si$_3$N$_4$) microring resonator (MRR)~\cite{Lu2021,nitiss_optically_2022} to translate a telecom (1560 nm) comb to 780 nm and injection-locked a FP diode to the translated comb. The ability to OIL with low-power combs allowed us to expand the effective wavelength range over which the SHG combs are usable, to include wavelengths where the SHG process produces $<5$~nW of optical power.

\section{Motivation}

Generating EO combs that enable high SNR measurements across a broad range of wavelengths involves several considerations to meet the wide range of potential applications shown in Fig.~\ref{Fig1}(a). Typically, a continuous-wave laser source and EO modulator are needed for each wavelength band of interest. While the linear EO effect in commonly used materials like lithium niobate is inherently broadband, accompanying effects such as photorefraction can result in pronounced power handling limitations at short wavelengths~\cite{damage_Hall1985}, necessitating studies of alternate materials~\cite{Kabessa2015}. Thus, even if low-noise laser sources with higher power are available (Fig.~\ref{Fig1}(a)-\textcircled{\small{1}}), the generation of EO combs with adequate power can be challenging (Fig.~\ref{Fig1}(a)-\textcircled{\small{2}}). A potential solution is to amplify a relatively low-power EO comb. While this is straightforward at some wavelengths (e.g., in the telecom where low-noise erbium-doped fiber amplifiers are available), it is far more difficult or not presently possible at many other wavelengths. Spectral translation of EO combs from technologically established wavelengths using either bulk~\cite{ns_res_Long2023, m_IR_DC_Yan2017, LuoComb, JerezComb} or on-chip nonlinear processes~\cite{Long2024} is another approach that bypasses the need for lasers and EO modulators working at different wavelengths, but places a premium on high conversion efficiency and output power, which can be challenging to achieve (Fig.~\ref{Fig1}(a)-\textcircled{\small{3}}). Finally, we note that certain spectroscopy targets, such as quantum or biological systems, may impose power limitations to avoid saturation or specimen damage (Fig.~\ref{Fig1}(a)-\textcircled{\small{4}}). In this case, the low-power EO comb which probed the sample can be effectively amplified prior to detection.

\begin{figure*}[!t]
\centering\includegraphics[width=0.75\linewidth]{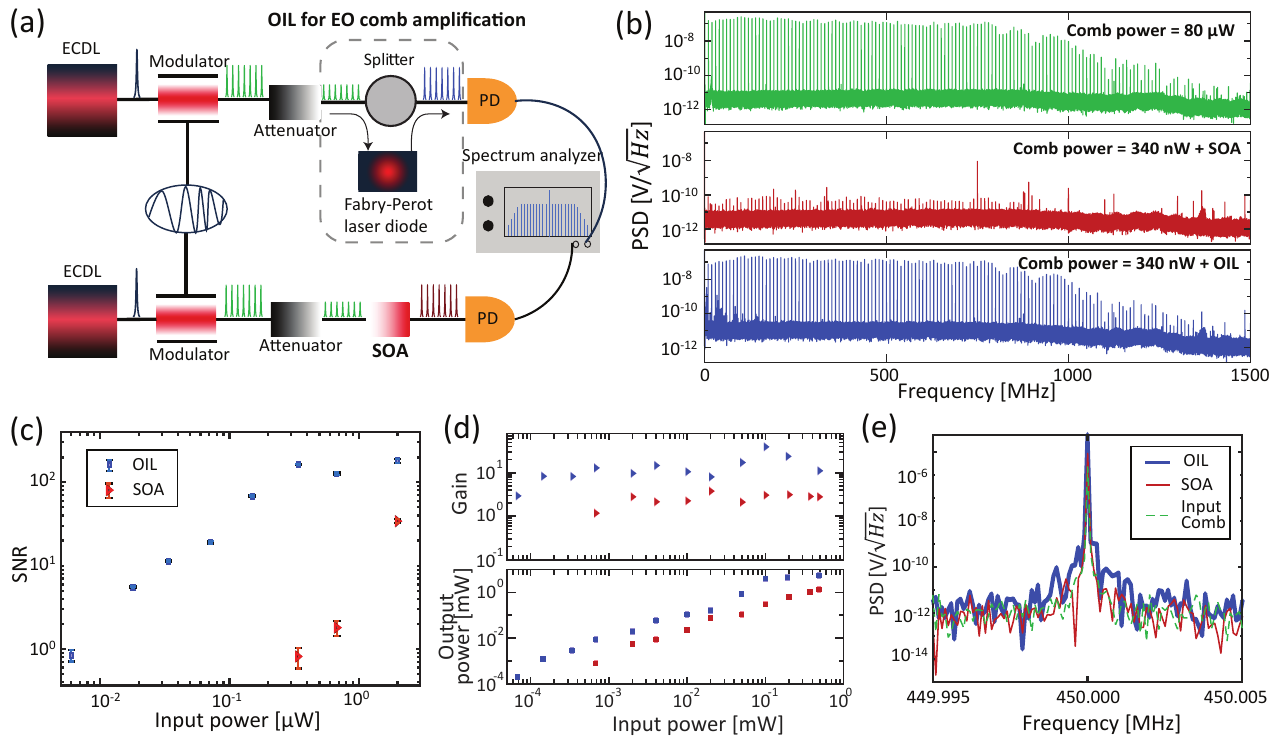}
\caption{ \textbf{Comparison of EO comb injection locking and semiconductor optical amplification.} \textbf{(a)} Setup schematic: A 780~nm external-cavity diode laser (ECDL) is phase-modulated by a frequency-chirped waveform in order to generate an EO comb. The EO comb is then sent through a variable optical attenuator for power control, and either routed to a Fabry-Perot (FP) laser diode (top) or through a semiconductor optical amplifier (SOA, bottom). The output from the FP laser diode or the SOA is measured with a fast photodetector (PD) and then analyzed using an electrical spectrum analyzer. \textbf{(b)} Recovery of EO comb SNR through OIL.  The top spectrum (green) shows an original high SNR EO comb at 80~$\mu$W of total comb power.  We subsequently attenuate that comb to a power of~340 nW, significantly reducing its SNR, before attempting to improve through an SOA (middle panel, red) and OIL (bottom panel, blue). The x-axis frequency here and in subsequent figures is with respect to the EO comb carrier (in this case, near 780 nm). \textbf{(c)} SNR of the amplified comb (red triangles) via the SOA and injection-locked FP diode (blue squares) as a function of input comb power. The displayed uncertainties are the standard deviation of repeated measurements made of the comb tooth amplitude and noise level. \textbf{(d)} Output power (squares, bottom graph) and gain (triangles, top graph) of the SOA-amplified comb (red) and injection-locked FP diode (blue). The uncertainty in the power measurement is 3~$\%$, represents a one standard deviation value, and is smaller than the marker size. \textbf{(e)} Comparison of an individual comb tooth linewidth between the amplified (red), locked FP diode (blue), and original input comb (green dashed) cases, for an input EO comb power of 80~$\mu$W. The line shapes for the amplified and OIL cases closely match that of the input EO comb.} 
\label{Fig2}
\end{figure*}

OIL of a high-power (FP) diode laser has long been used to effectively increase the available power while retaining the characteristics of a low-noise laser~\cite{Liu2020}. In recent years, advances in low-loss photonic integrated circuits (PICs), together with the widespread commercial availability of low-cost FP laser diodes, have led to the creation of hybrid-integrated compact and low-noise laser sources across the visible and short near-infrared~\cite{corato-zanarella_widely_2023,lu_emerging_2024}. Here, we explore another aspect of OIL, enabling the effective amplification of an entire EO comb (Fig.~\ref{Fig1}(b)). 

While OIL has been used with seed light from optical frequency combs in numerous previous studies~\cite{Liu2020}, including work using EO combs~\cite{fukushima_optoelectronic_2003,xu_wideband_2021,skehan_widely_2021}, these studies have focused on amplification of individual comb teeth or amplification via diode laser arrays~\cite{zhang_high-coherence_2024}. Multi-wavelength injection locking has also been studied~\cite{chen_multi-wavelength_2022}, where up to 30 channels have been self-injection locked using an external cavity, across a total bandwidth on the order of 1.5~THz. Here, we study OIL in a different regime (Fig~\ref{Fig1}(b), the left dashed box), whereby up to two million comb teeth are simultaneously amplified by a single FP laser diode. Our input combs are generated by phase-modulation of a continuous wave laser, and our experiments thus probe dynamic OIL in which the seed laser phase is varied.  
We investigate two regimes, one where we use a chirped comb whose entire bandwidth fits within the FP diode's longitudinal mode spacing, and one where we use an over-driven comb whose bandwidth spans over 400~GHz and several longitudinal modes of the diode. We then further show that this OIL technique can overcome inefficiency challenges associated with chip-integrated frequency converters (Fig~\ref{Fig1}(b), the right dashed box).

\section{Injection locking a Fabry-Perot laser diode to an EO comb}

\begin{figure*}[t!]
\centering\includegraphics[width=0.75\linewidth]{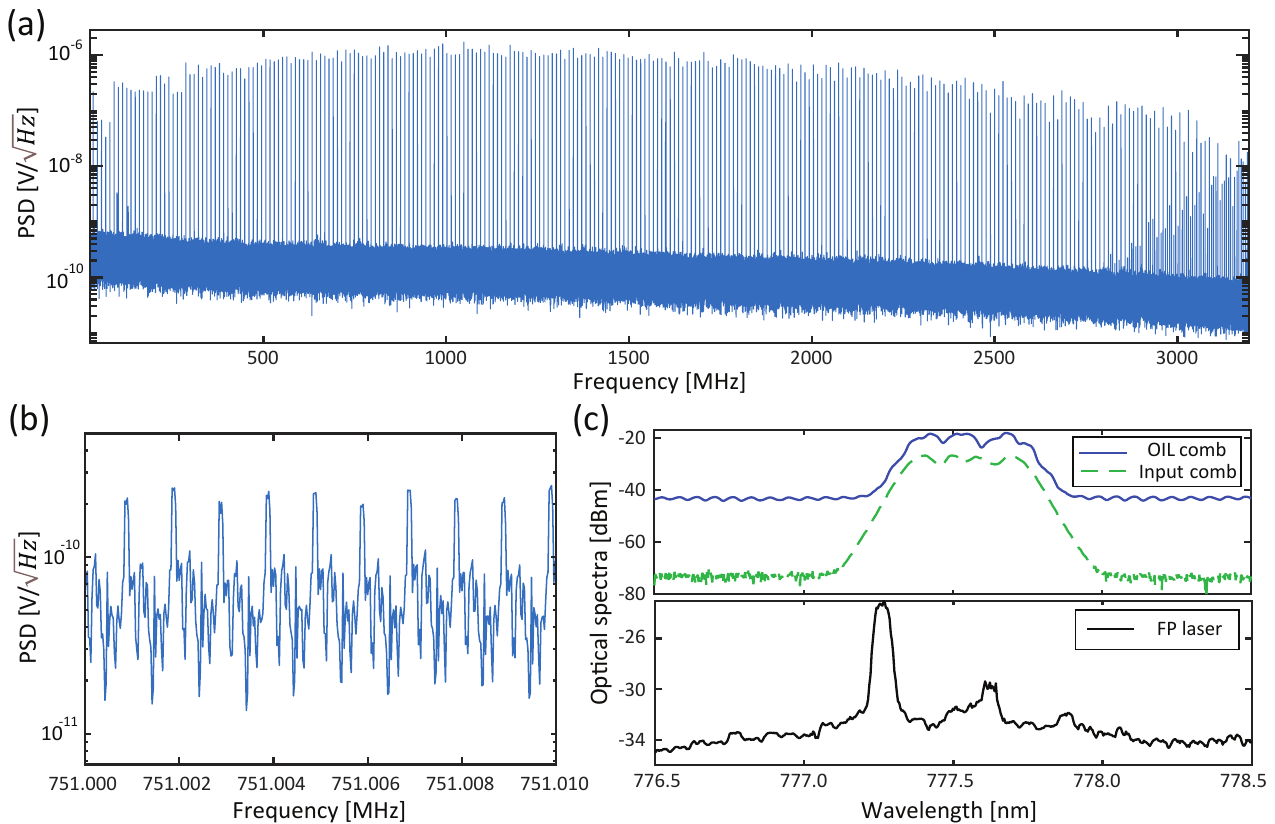}
\caption{ \textbf{Injection-locked FP diode comb tunability.} \textbf{(a)} 6400~MHz wide comb at an 11.52 MHz spacing, for a total comb injection power of 10~$\mu$W. Aliasing is observed around 3~GHz due to imperfections in the chirped RF signal generating the comb. \textbf{(b)} Zoom-in of a 2~GHz wide comb with 1~kHz spacing, which contains a total of two million comb teeth. Only 8~$\mu$W of total injected power was used here. \textbf{(c)} In the top panel we present OIL of the FP laser locked to an EO comb with a 16~GHz spacing across $\approx$400~GHz bandwidth (blue) and the input comb (dashed green). In the bottom panel we present the free-running FP laser. The FP gain bandwidth exceeds the displayed range and goes from $\approx$775 nm to 780~nm. These measurements were captured using an optical spectrum analyzer (OSA)  with a resolution of 4~GHz.
}
\label{Fig3}
\end{figure*}

We utilized the setup depicted in Fig.~\ref{Fig2}(a) to demonstrate a FP OIL to a frequency-agile EO comb and to provide a direct comparison to the SOA approach. The output of a 780 nm external-cavity diode laser was passed through an EO phase modulator driven by a frequency-chirped waveform, resulting in an EO comb whose tooth spacing is given by the waveform's repetition rate and bandwidth is given by twice the chirp range~\cite{Long2016, long_electro-optic_2019}. For the present measurements repetition rates (and therefore the comb tooth spacing) were varied between 1 kHz and 11.52 MHz and the comb span was between 2 GHz and 6.4 GHz. The comb power was controlled by a variable optical attenuator, after which it was sent through a 50:50 splitter and then to a 780~nm FP diode. One of the FP diode modes was temperature-tuned to the injected laser's wavelength for OIL to the EO comb. The output of the FP diode was collected via the other port of the splitter and sent to a high-bandwidth detector and electrical spectrum analyzer. For comparison, we also amplified the attenuated comb using an SOA and analyzed the output similarly. 

We compared the OIL and SOA results for a range of EO comb input powers and found significant benefits to the OIL approach. In Fig.~\ref{Fig2}(b), we start by displaying a high SNR EO comb obtained at an overall comb power of 80 $\mu$W, and with a 10~MHz spacing and 10~dB bandwidth of 2000~MHz.  We then attenuate this comb to a much lower power of 340~nW, significantly degrading its SNR as the comb tooth power goes below our detector noise floor.  At this total comb power of 340~nW, we consider amplification via the SOA (red) and OIL (blue). While a clear comb is present in the OIL case, the SOA produces a very low-quality comb with not all comb teeth visible above the noise level. Although 340~nW is thus near the minimum input power we can successfully amplify with the SOA, with OIL we can work with substantially lower powers. For a more quantitative assessment, we define the SNR as the average amplitude of the comb teeth divided by the noise level, and we compare the SNR for the OIL and SOA cases for various input powers in Fig.~\ref{Fig2}(c). With injection locking, we observe SNR $ >10 $ for inputs as low as 34~nW, whereas with the SOA, the SNR was only  $ >1 $ for input powers $ >1 $~$\mu$W. In Fig.~\ref{Fig2}(d), we present the output power and gain for both the OIL and SOA cases as a function of input comb power. For the optimized OIL and SOA parameters (temperature and current), OIL also produced higher output power and gain. In Fig.~\ref{Fig2}(e), we plot a single tooth of an optical comb produced by OIL and the SOA and observe that they are comparable in linewidth, thus indicating that the enhanced power does not incur additional phase noise.

\begin{figure*}[t!]
\centering\includegraphics[width=0.75\linewidth]{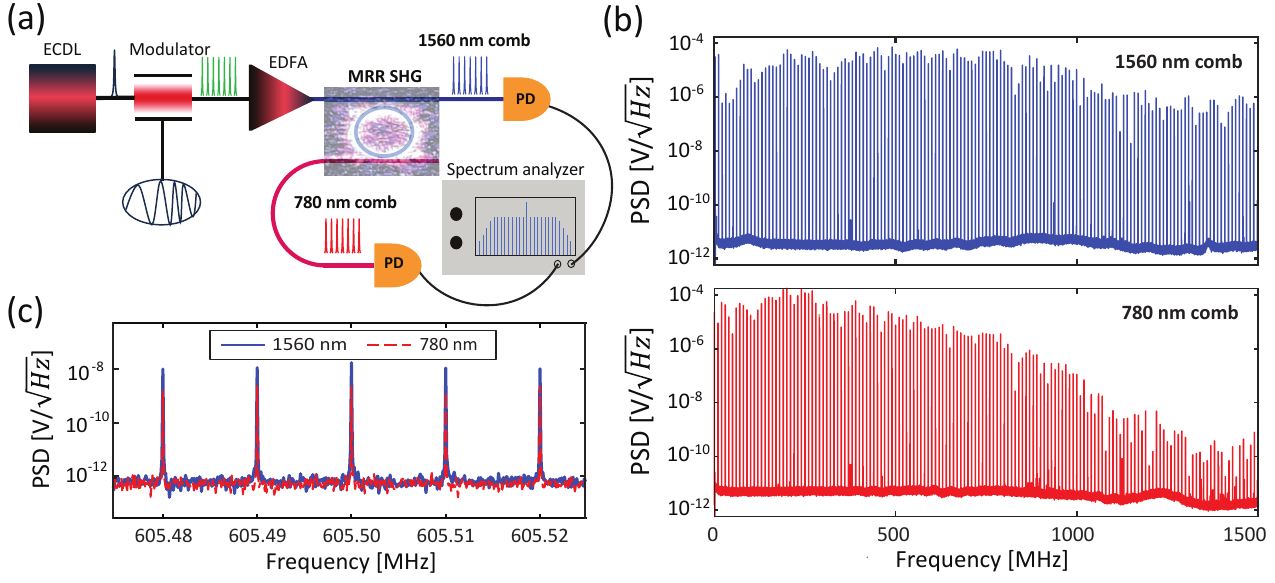}
\caption{ \textbf{Spectral translation of an EO comb using nanophotonic SHG/SFG.} 
\textbf{(a)} Setup schematic: A telecom laser is phase-modulated with a frequency-chirped waveform to generate an EO comb. This comb is then amplified using an erbium-doped fiber amplifier (EDFA). The comb is then routed to a Si$_3$N$_4$ microring resonator (MRR) where the SHG signal is generated. Both the telecom comb that is not coupled into the MRR and the SHG output (extracted from the MRR through a separate bus waveguide) are measured using fast photodetectors (PDs). \textbf{(b)} 1560~nm comb transmitted through the top bus waveguide (blue). Spectrally-translated 780~nm comb extracted by the bottom bus waveguide (red). \textbf{(c)} High resolution combs (10~kHz tooth spacing) at 1560~nm  and 780~nm. The 780~nm comb having the same spacing as the 1560~nm comb, suggesting that the underlying nonlinear conversion processes include both second harmonic generation and sum-frequency generation.}
\label{Fig4}
\end{figure*}

From the above, we find that for EO combs, OIL provides clear SNR and output power benefits in comparison to semiconductor optical amplification for input powers between 6~nW and 2~$\mu$W. Given that EO comb applications benefit from flexibility in the comb parameters, we have locked the FP laser diode to combs with various bandwidths and spacing. For example, in Fig.~\ref{Fig3}(a), we show an injection-locked comb with 11.52 MHz spacing and a total bandwidth of 6400~MHz; in comparison to Fig.~\ref{Fig2}(c), here the overall comb bandwidth is increased by a factor of six. In Fig.~\ref{Fig3}(b), we reduce the comb tooth spacing by four orders of magnitude, to 1~kHz, and show that this fine-tooth comb, suitable for high-resolution spectroscopy, can also successfully serve as the seed for OIL. While the figure shows a zoom-in of a 10~kHz portion of the comb to highlight the comb tooth spacing, we note that the total comb span is 2~GHz, so that there are 2$\times10^6$ teeth that injection lock the FP diode. 

Thus far, we have focused on EO combs, created by chirped phase modulation, which address applications where spectral resolution in the $\approx$1~kHz to $\approx$10~MHz range is needed, and total comb bandwidths $\lesssim5$~GHz are sufficient. However, because the FP diode we used has a typical linewidth of $\approx$0.5~THz (Fig.~\ref{Fig3}(c), black), and is comprised of several longitudinal modes, we can consider injection locking using EO combs with substantially larger bandwidths. We tested such broadband operation by overdriving our modulator (i.e., at multiple times its half-wave voltage) at a frequency of 16~GHz and injecting it into the FP diode. In the top panel of Fig.~\ref{Fig3}(c), we plot the spectrum of the FP laser locked to this 16~GHz repetition rate EO comb (blue) and compare it to the input comb (green), observing an effective amplification of the comb by $\approx10$~dB for a comb consisting of more than 25 teeth across $\approx$~400~GHz of bandwidth. Moreover, in the bottom panel of Fig.~\ref{Fig3}(c), we plot the spectrum of the free-running FP laser (black). Its gain bandwidth exceeds the displayed range, indicating injection locking to an even broader bandwidth comb should be possible.

\section{Spectral translation of an EO comb using SHG}

Our study has thus far focused on OIL of a 780~nm FP laser diode using an EO comb generated with a narrow linewidth seed laser and modulator at the same 780~nm wavelength. However, as high-quality (e.g., narrow linewidth) sources, amplifiers, and modulators are more abundant at telecom wavelengths, it would be beneficial to translate telecom EO combs to 780~nm using second harmonic generation (SHG). The ability to spectrally translate EO combs using other parametric nonlinear processes has been recently demonstrated~\cite{ns_res_Long2023,Long2024}. Apart from the SHG process providing a larger relative frequency shift than the processes shown in the earlier works, here we show that a key limitation associated with such spectral translation -- limited conversion efficiency and output power -- can be overcome with OIL. We investigate this claim by injection locking a 780 nm FP laser to a 780~nm EO comb produced through SHG of a telecom (1560~nm) comb. 

\begin{figure*}[t!]
\centering\includegraphics[width=0.75\linewidth]{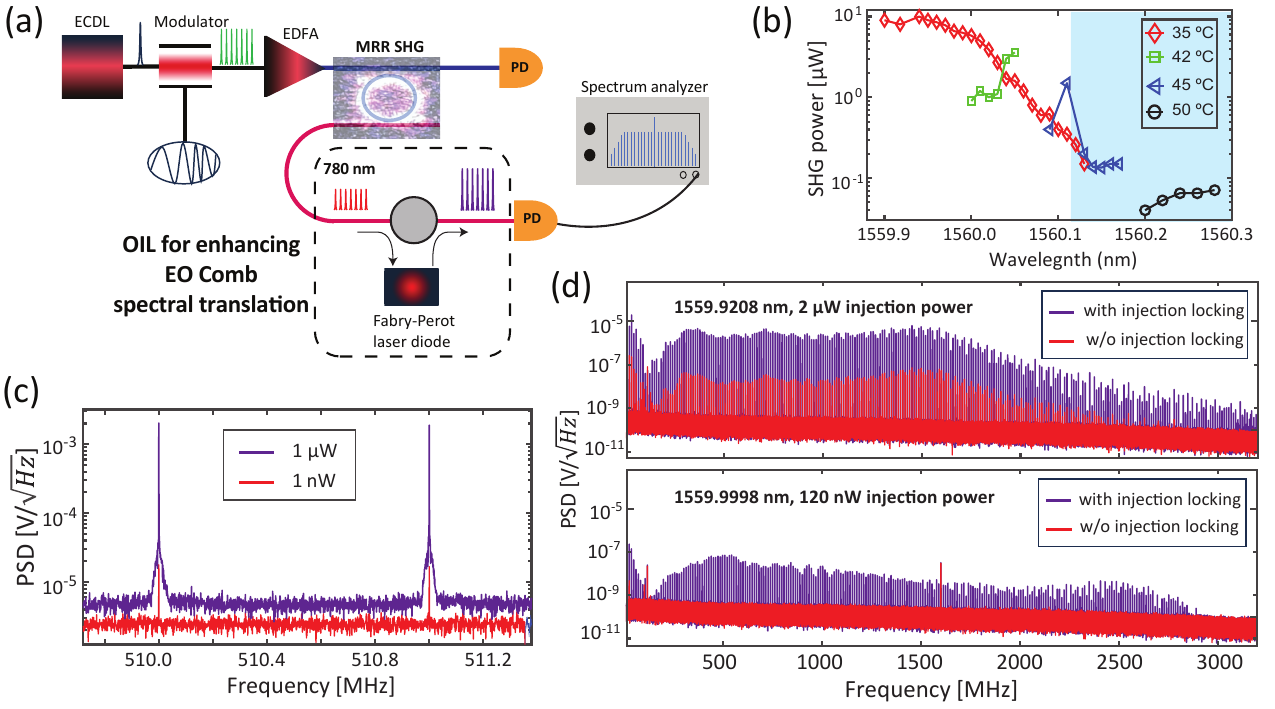}
\caption{ \textbf{Expanding the usable range of EO comb spectral translation using injection locking.}  
\textbf{a} Setup schematic for injection locking using the SHG EO comb: A telecom laser is modulated to generate an EO comb. The comb is amplified by an EDFA and fed into an MRR. Through SHG/SFG, a 780~nm comb is generated and injected into an FP laser diode through a splitter. The FP diode output is measured via a fast photodetector and electronic spectrum analyzer. \textbf{(b)} SHG output as a function of the input telecom wavelength and the temperature. The shaded blue area shows the region of operation where we are able to observe a spectrally translated comb using injection locking but not using an SOA. \textbf{(c)} FP diode laser output after being injection locked to the SHG/SFG comb attenuated to a total power of $\approx$~1~$\mu$W (purple) and $\approx$~1~nW (red). \textbf{(d)} SHG/SFG comb (red) and FP diode laser output after being injection locked to the SHG/SFG comb (purple) at two different wavelengths and injection powers. The uncertainty in the listed wavelength in (d) is 0.1~pm, one standard deviation associated with the accuracy of the wavemeter. In the bottom panel, the SHG/SFG comb power is too low to resolve without injection locking.}
\label{Fig5}
\end{figure*}

First, we show our ability to translate an EO comb from the telecom to 780~nm, using the setup depicted in Fig.~\ref{Fig4}(a). A 1560 nm EO comb is generated by phase modulation of a cw telecom external cavity diode laser via a frequency-chirped waveform. The comb is then amplified by an erbium-doped fiber amplifier (EDFA) and fed to a Si$_3$N$_4$ microring resonator (MRR). When there is momentum and energy matching between the telecom and 780~nm modes in the MRR, SHG can be realized through the DC-Kerr effect, with the DC field generated by the coherent photogalvanic effect~\cite{Lu2021,nitiss_optically_2022}. We used an MRR fabricated in the same batch as the one published in Ref.~\cite{Lu2021}, and further explanation of the MRR fabrication process and mode matching verification can be found in that work. Of specific relevance here is that the MRR is coupled using two bus waveguides, with the top waveguide in Fig.~\ref{Fig4}(a) dedicated to in-coupling 1560~nm light, and the bottom waveguide dedicated to out-coupling the generated 780~nm SHG light. In Fig.~\ref{Fig4}(b), we plot the transmitted telecom comb (i.e., the portion of the 1560~nm EO comb that is not coupled into the ring) and also the extracted frequency-converted 780~nm comb. We note two particularly important points regarding the frequency-converted comb. First, there is an overall bandwidth reduction compared to the telecom comb. This is a result of the filtering effect of the 1560~nm cavity mode, which admits only a portion of the input telecom comb, as well as the higher quality factor of the MRR at 780~nm in comparison to 1560~nm. Second, after zooming in on the combs (Fig.~\ref{Fig4}(c)), we see that the comb teeth have similar linewidth, as can be expected from a coherent process, and that they have the same spacing. The latter point indicates that each spectrally translated comb tooth is not generated exclusively by SHG from a single telecom comb tooth, but rather by a combination of SHG and SFG.

\section{Injection locking of low power spectrally translated combs}
In this section, we further implement optical injection locking in the spectral translation of an EO comb, as shown in Fig.~\ref{Fig5}(a). This setup is similar to that in Fig.~\ref{Fig4}(a) yet with an additional optical injection locking unit (dashed box). While the photo-induced SHG process in MRRs can be realized with high efficiency~\cite{Lu2021}, there are a number of situations for which the efficiency and/or output power are limited.  For example, we use a doubly resonant process for SHG, so that maximizing efficiency relies on obtaining a precise harmonic relationship between the fundamental (1560~nm band) and second harmonic (780~nm band) cavity modes. Typically, this will occur for a specific cavity temperature, and it can be challenging if one additionally requires the second harmonic light to be at a precise wavelength. For example, as we heated the ring and red-shifted the modes, the efficiency of the SHG process was reduced (Fig.~\ref{Fig5}(b)), which we attribute to the modes' increased misalignment (i.e., a departure from being strictly harmonic). 

In particular, temperature tuning shifted the SHG light by 0.22~nm but reduced the SHG power from $\approx$~5~$\mu$W to $\approx$~5~nW. To rectify this and recover a usable comb power near 780~nm, we then injection lock a 780~nm FP laser diode using the SHG light at various wavelengths and powers (setup in Fig.~\ref{Fig5}(a)). First, we examined the lowest SHG/SFG comb power to which we can lock the FP diode. In Fig.~\ref{Fig5}(c), we plot the partial spectrum of this OIL FP laser, and show that we can lock the diode using a comb with a total power of $\approx~$1~nW and $<1$~pW of power per single tooth. In Fig.~\ref{Fig5}(d), we show two cases of the spectrally-translated EO comb at two different wavelengths, with and without injection locking, respectively, in red and blue. We see an improvement in the SNR with injection locking (the top panel), and in the case where the total SHG/SFG power is $\approx$~120~nW (the bottom panel), we are unable to resolve a 780~nm comb without injection locking. This clearly shows the potential of using injection locking to expand the usable operating wavelengths of spectrally translated combs.

%
%
\section{Discussion}
In summary, we have shown that we can inject a low-power EO comb into a Fabry-Perot laser diode and lock it, generating a high SNR EO comb at very low comb powers that cannot be effectively increased through direct semiconductor optical amplification. As a demonstration of the usefulness of this method in overcoming low SNR, we expand the operation bandwidth of combs translated by integrated nonlinear devices.  Though in this work, we focus on EO combs in the 1560~nm and 780~nm spectral regions, the aforementioned availability of Fabry-Perot laser diodes across a broad range of wavelengths, as well as continued progress in expanding wavelength access through nonlinear nanophotonics, suggests that our approach can be directly applied to many different EO comb spectroscopy applications. We believe this method can be further expanded to the visible and mid-IR and allow for the generation of high-power, narrower linewidth EO combs using Fabry-Perot diodes, a cavity-stabilized telecom laser, and cascaded integrated nonlinear processes.

\section{Acknowledgements} The authors acknowledge support from the NIST-on-a-chip program and helpful comments from Daniel Pimbi and Benjamin Reschovsky. A provisional patent has been filed on which several of the authors are listed as inventors (R.Z., A.C., D.A.L., K.S.).


\end{document}